\documentclass[pre,amsmath,amssymb,10pt,nofootinbib]{revtex4}

\usepackage{url}
\usepackage{hyperref}
\hypersetup{
    colorlinks,
    citecolor=blue,
    filecolor=black,
    linkcolor=blue,
    urlcolor=blue
}
\usepackage{mathrsfs}
\usepackage{verbatim}
\usepackage{graphicx}	
\usepackage{dcolumn}	
\usepackage{bm}		
\usepackage{color}
\usepackage[cal=boondoxo]{mathalfa}
\usepackage{palatino}
\usepackage{dsfont}
\usepackage{cancel}
\usepackage{xcolor}

\DeclareFontFamily{U}{euc}{}
\DeclareFontShape{U}{euc}{m}{n}{<-6>eurm5<6-8>eurm7<8->eurm10}{}
\DeclareSymbolFont{AMSc}{U}{euc}{m}{n}
\DeclareMathSymbol{\umu}{\mathord}{AMSc}{"16}

\DeclareSymbolFont{AMSb}{U}{msb}{m}{n}
\DeclareMathSymbol{\N}{\mathbin}{AMSb}{"4E}
\DeclareMathSymbol{\Z}{\mathbin}{AMSb}{"5A}
\DeclareMathSymbol{\R}{\mathbin}{AMSb}{"52}
\DeclareMathSymbol{\Q}{\mathbin}{AMSb}{"51}
\DeclareMathSymbol{\I}{\mathbin}{AMSb}{"49}
\DeclareMathSymbol{\C}{\mathbin}{AMSb}{"43}

\begin{document}

\title{The development of an information criterion for Change-Point Analysis.}
\author{Paul A. Wiggins}

\author{Colin H. LaMont}
\affiliation{Departments of Physics, Bioengineering and Microbiology, University of Washington, Box 351560.\\ 3910 15th Avenue Northeast, Seattle, WA 98195, USA}

\email{pwiggins@uw.edu}\homepage{http://mtshasta.phys.washington.edu/}
%


\begin{abstract}
Change-point analysis is a flexible and computationally tractable tool for the analysis of times series data from systems that transition between discrete states and whose observables are corrupted by noise. The change-point algorithm is used to identify the time indices (change points) at which the system transitions between these discrete states. We present a unified information-based approach to testing for the existence of change points. This new approach reconciles two previously disparate approaches to Change-Point Analysis (frequentist and information-based) for testing transitions between states. The resulting method is statistically principled, parameter and prior free and widely applicable to a wide range of change-point problems.

\end{abstract}


\maketitle


\section{Introduction}

The problem of determining the  true state of a system that transitions between discrete states and whose observables are corrupted by noise is a canonical problem in statistics with a long history (e.g. \cite{Little:2011uq}). The approach we discuss in this paper is called Change-Point Analysis and was first proposed by E. S. Page in the mid 1950s \cite{Page1955,Page1957}.  
 Since its inception, Change-Point Analysis has been used in a great number of contexts and is regularly re-invented in fields ranging from geology to biophysics \cite{chen2007,Little:2011uq,Little:2011kx}.  
 
The primary goal of this paper is to develop a new information-based approach to Change-Point Analysis which simplifies its application in problems, including those where a specific change-point statistics have not been computed. 
 We approach  Change-Point Analysis from the perspective of {\it Model Selection} and {\it Information Theory}. Akaike pioneered a powerful approach to Model Selection by the minimization of the Kullback-Leibler  Divergence \cite{Kullback1951a}, a measure of information loss by approximating the true process with a model \cite{akaike1773,BurnhamBook}. He demonstrated that  two key principles of modeling, predictivity and parsimony, were in fact conceptually and mathematically linked (e.g. \cite{BurnhamBook}). In short, the addition of superfluous parameters to a model, reducing parsimony, results  in information loss, reducing predictivity  (e.g. \cite{BurnhamBook}). Akaike derived an unbiased estimator for information loss, the Akaike Information Criterion (AIC), which proved to be at once exceptionally tractable and widely applicable.   

Unfortunately Akaike's approach is limited to regular models \cite{watanabe2009}. Change-Point Analysis and many other applications are singular. These models contain unidentifiable parameters with nearly zero Fisher Information, which greatly increase the complexity of the model and lead to the catastrophic failure of AIC to estimate information loss. The subject of this paper is  the  implementation of information-based model selection in the context of Change-Point Analysis. We have recently proposed a Frequentist Information Criterion (FIC) applicable even in the context of singular models. Using FIC and an approximation analogous to that used by Akaike to derive AIC, we develop a model criterion that accounts for the unidentifiability of the change-point indices. Importantly, this criterion does not depend on the detailed form of the model for the individual states but only on the number of model parameters, in close analogy with AIC. Therefore we expect this result to be widely applicable anywhere the change-point algorithm is applied. 

Frequentist statistical tests have already been defined for a number of canonical change-point problems. It is therefore interesting to  examen the relation between this approach and our newly-derived information-based approach. We find the approaches are fundamentally related. The information-based approach can be understood to provide an predictively-optimal confidence level for a generalized ratio test. The Bayesian Information Criterion (BIC) has also been used in the context of Change-Point Analysis. We find very significant differences between our results and the BIC complexity that suggest that BIC is not suitable for application to change-point analysis since it can lead to either over or under segmentation of the data, depending on the specific context.


\section{Preliminaries}

\begin{figure}
  \centering
    \includegraphics[width=1.0\textwidth]{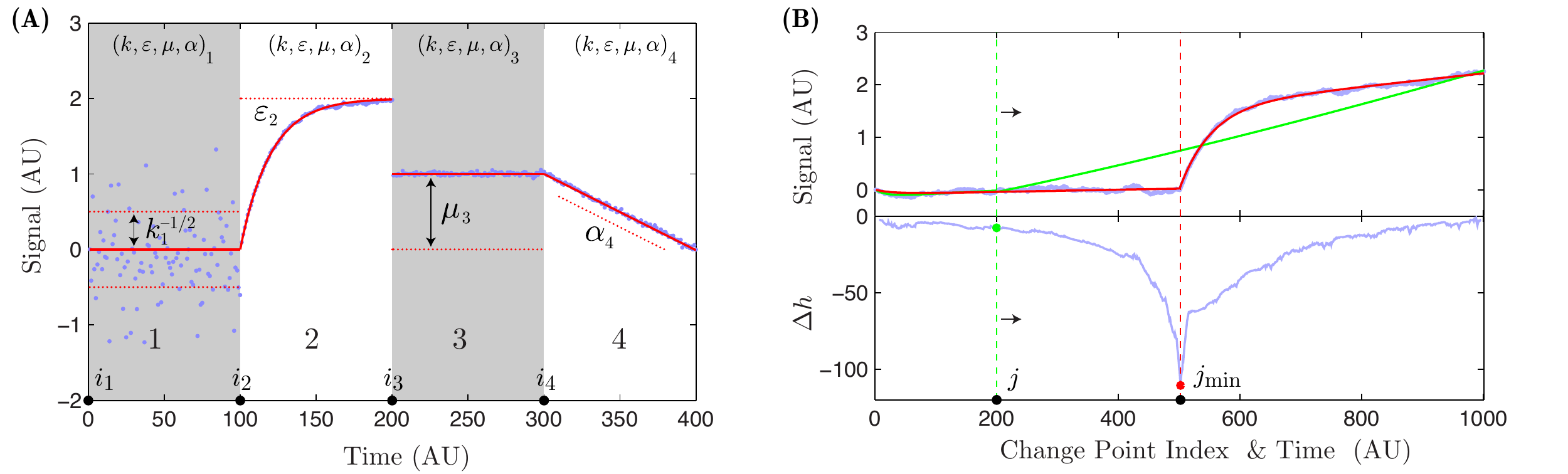}
      \caption{{\bf Panel A: State model schematic.} A state model for biophysical applications is parameterized by four model parameters that are written as the vector ${\bm \theta}\equiv(k,\varepsilon,\mu,\alpha)$. Above we schematically illustrate the role of each parameter in shaping the signal. 
          \label{schematicFig} {\bf Panel B: Schematic of  binary segmentation.} To segment a partition, the information change due to placing a change point at each time index along the x axis is considered. (The dashed red and green lines represent possible change points.) For each change-point index, a minimum-information fit is performed on the two resulting data partitions (top panel, blue dots), resulting in the solid curves (top panel, red and green for the respective change-point indices). For each change-point index, an information change is computed (bottom panel). The change point is placed at the time index that minimizes the information change (red dashed line).
          \label{schematic_changepointFig}}
\end{figure}



\medskip
\noindent
We introduce the following notation for a signal: a set of $N$ ordered observations from a one-dimensional stochastic process\footnote{When $X$ appears in upper case, it should be understood as a random variable whereas it is a normal variable when it appears in lower case. If we need  a statistically independent set of variables of equal size, we will use the random variables $Y^N$, which have identical properties to the $X^N$.}:
\begin{equation}
X^N \equiv (X_1,X_2,...,X_N)\sim p(\cdot), 
\end{equation}
where the observation index is often but not exclusively temporal and the probability distribution for the stochastic process is represented as $p$. We shall represent the probability distribution for the model $\cal M$ as:
\begin{equation}
q(X^N|{\cal M}),
\end{equation} 
where these is no guarantee that true distribution is a member of the model family. 

\medskip
\noindent
{\bf Information and cross entropy.} The coding information for signal $X^N$ given model $\cal M$ is:
\begin{equation}
h(X^N|{\cal M}) \equiv -\log q(X^N|{\cal M}), \label{Eqn:info}
\end{equation}
and the cross entropy for the signal (average coding information) is:
\begin{equation}
H^N({\cal M}) \equiv \underset{p(\cdot)}{{\mathbb E}_X} h(X^N|{\cal M}), \label{Eqn:CrossEnt}
\end{equation}
where the expectation over the signal $X^N$ is understood to be taken over the true distribution $p$.

\medskip
\noindent
{\bf The Change-Point Model.} We define a model for the signal corresponding to a system transitioning between a set of discrete states. We define the discrete time index corresponding to the start of the $I$th state $i_I$. This index is called a {\it change point}. The model parameters describing the signal in the $I$th interval  are ${\bm \theta}_I$.  Together these two sets of parameters ($i_I$ and ${\bm \theta}_I$) parameterize the model ${\cal \Theta}$. The model parameterization for the signal (including multiple states) can then be written explicitly:
\begin{eqnarray}
{\cal M}^n &=& \left( \begin{array}{cccc}
1 & i_2 & \hdots & i_{n}\\
{\bm \theta}_1  & {\bm \theta}_2 & \hdots & {\bm \theta}_{n} \\  
\end{array}
 \right),
\end{eqnarray}
where  $n$ is the number of states or change points. A schematic example of a change-point mode for a biophysical signal is shown in Figure~\ref{schematic_changepointFig}.
The two sets of parameters (${\bm \theta}_I$ and $i_I$) are fundamentally different. We shall assume that the state model is regular: i.e.~the parameters ${\bm \theta}_I$ have non-zero Fisher information \cite{FICshort}. By contrast, the change-point indices $i_I$ are discrete and typically non-harmonic parameters. For instance, consider a true model $p=q$ where $\bm \theta_1 = \bm \theta_2$. In this scenario the cross entropy will be independent of $i_2$ as long as $i_2\in (i_1,i_3)$. The Fisher information corresponding to $i_2$ is threfore zero. These properties  have important consequences for model selection \cite{FICshort}.

\medskip
\noindent
{\bf Determination of model parameters.} Fitting the change-point model is performed in two coupled steps. Given a set of change-point indices ${\bm i }^n \equiv (i_1,...,i_n)$, the maximum likelihood estimators (MLE) of the state model parameters ${\bm \Theta}^n\equiv ({\bm \theta}_1,...,{\bm \theta}_n)$ are defined:
\begin{equation}
\hat{ \bm \Theta}^n_X = \arg \min_{{\bm \Theta}^n}  h(X^N|{\cal M}^n). 
\end{equation} 
The determination of the change-point indices ${\bm i}^n$ is a nontrivial problem since not only are the change-point indices unknown, but even the number of transitions ($n$) is unknown. 

\medskip
\noindent
{\bf Binary Segmentation Algorithm.} To determine the change-point indices, we will use a binary-segmentation algorithm that has been the subject of extensive study (e.g see the references in \cite{chen2007}). In the global algorithm, we initialize the algorithm with a single change point $i_1=1.$ The data is sequentially divided into partitions by binary segmentation. Every segmentation is {\it greedy}: i.e. we choose the change point on the interval $(1,N)$ that minimizes the information in that given step, without any guarantee that this is the optimum choice over multiple segmentations. The family of models generated by successive rounds of segmentation are said to be {\it nested} since successive changes points are added without altering the time indices of existing change points.  Therefore, the previous model is always a special case of the new model.
The binary segmentation process is shown schematically in Fig.~\ref{schematic_changepointFig}, Panel B. In each step, after the optimum index for segmentation is identified, we statistically test the change in information  (due to segmentation) to determine whether the new states are statistically supported.  The change-point determined by binary segmentation determine the change-points in the MLE model $\hat{\cal M}^n$.
The local binary-segmentation algorithm differs from the global algorithm only in that we consider the binary segmentation of each partition of the data independently. The algorithms as described explicitly in the supplement.

\medskip
\noindent
{\bf Information-based model selection.} The model that minimizes the cross entropy (Eqn.~\ref{Eqn:CrossEnt}) is the most predictive model. Unfortunately, the cross entropy cannot be computed since the expectation cannot be taken with respect to the true but unknown probability distribution $p$ in Eqn.~\ref{Eqn:CrossEnt}. The natural estimator of the cross entropy is the information (Eqn.~\ref{Eqn:info}), but this estimator is biased from below: Due to the phenomena of over-fitting, added model parameters always reduce the information (or equivalently the training error) even as the predictivity of the model is reduced by the addition of superfluous parameters. We must therefore construct an unbiased estimator of the cross entropy which we call the {\it information criterion}:
\begin{equation}
{\rm IC}(X^N,n) \equiv h(X^N|\hat{\cal M}^n_X) + {\cal K}(n),
\end{equation}
where ${\cal K}$ is the complexity of the model which is defined as the bias in the information as an estimator of cross-entropy:
\begin{equation}
 {\cal K}(n) \equiv \underset{p}{{\mathbb E}_{X,Y}} \left\{h(Y^N|\hat{\cal M}^n_X)-h(X^N|\hat{\cal M}^n_X)\right\},
\end{equation}
where the expectations are taken with respect to the true distribution $p$ and $X^N$ and $Y^N$ are independent signals.
For a regular model in the asymptotic limit, the complexity is equal to the number of model parameters and the information criterion is equal to AIC. In the context of singular models, a more generally applicable approach must be used to approximate the complexity.

\medskip
\noindent
{\bf Frequentist Information Criterion.} The Frequentist Information Criterion (FIC) uses a more general approximation to estimate the model compleixty. Since the true distribution $p$ is unknown, we make a frequentist approximation, computing the complexity for the model ${\cal M}$ as a function of the true parameterization:
\begin{equation}
{\cal K}_{\rm FIC}( {\cal M}^n,n ) \equiv \underset{q(\cdot|{\cal M}^n)}{{\mathbb E}_{X,Y}} \left\{h(Y^N|\hat{\cal M}^n_X)-h(X^N|\hat{\cal M}^n_X)\right\},
\label{Eqn:FICComp} 
\end{equation}
and the corresponding information criterion is defined:
\begin{equation}
{\rm FIC}(X^N,n) \equiv h(X^N|\hat{\cal M}^n_X) + {\cal K}_{\rm FIC}(\hat{\cal M}^n_X,n),
\label{Eqn:defBias2} 
\end{equation}
where the complexity is evaluated at the MLE parameters $\hat{\cal M}^n_X$. The model that minimizes FIC has the smallest expected cross entropy.

\medskip
\noindent
{\bf Approximating the FIC complexity.} The direct computation of the FIC complexity (Eqn.~\ref{Eqn:FICComp}) appears daunting, but a tractable approximation allows the complexity to be estimated. The complexity difference between the models is:
\begin{eqnarray}
{\cal k}(n) &\equiv& {\cal K}_{\rm FIC}(n)-{\cal K}_{\rm FIC}(n-1), \label{Eqn:nestingpen}
\end{eqnarray}
which is called the nesting complexity.
An approximate piecewise expression can be computed as follows.
Let the observed change in the MLE information for the $n$th nesting be
\begin{equation}
\Delta h_n \equiv h(X^N|\hat{\cal M}^n_X)-h(X^N|\hat{\cal M}^{n-1}_X),
\end{equation}
where $n$ denotes the $n$th nesting of model ${\cal M}$. Consider two limiting cases: When the new parameters are identifiable, let the nesting complexity be given by ${\cal k}_+$ whereas when the new parameters are unidentifiable, let the nesting complexity be given by ${\cal k}_-$. When the new parameters are identifiable, the model is essentially regular therefore:
\begin{equation}
{\cal k}_+ = {\cal d},
\end{equation}
where ${\cal d}$ is the number of harmonic\footnote{ Harmonic parameters are parameter with sufficiently large Fisher information that they are not unidentifiable.} parameters added to the model in the nesting procedure, as predicted by AIC. 

To compute ${\cal k}_-$, we assume the unnested model is the true model and compute the complexity difference in Eqn.~\ref{Eqn:nestingpen}.
We then apply a piecewise approximation for evaluating the nesting complexity \cite{FICshort}:
\begin{equation} 
{\cal k}(n) \approx \begin{cases}
{\cal k}_-(n), & -\Delta h_{n} < {\cal k}_-(n)\\
{\cal k}_+(n), &{\rm otherwise}
\end{cases}.\label{Eqn:nestingcomplexity}
\end{equation} 
Since the nesting complexity represents complexity differences, the complexity can be summed:
 \begin{equation}
{\cal K}_{\rm FIC}(n) \equiv \sum_{j=1}^n {\cal k}(j), \label{Eqn:complexity}
\end{equation}
where the first term in the series, ${\cal k}(1)$ is computed using the AIC expression for the complexity.
An exact analytic description of the complexity remains an open question.


\section{An information criterion for chagne-point analysis}

\medskip
\noindent
{\bf Complexity of a state model.} As a first step towards computing the complexity for the change-point algorithm, we will compute the complexity for a signal with only a single state. It will be useful to break the information into the information per observation. Using the Markov property of the process, the information associated with the $i$th observation is:
\begin{equation}
h_i(X^N|{\bm \theta}) \equiv - \log q(X_i|X_{i-1}; {\bm \theta}). 
\end{equation}
For a stationary process, the average information per observation is constant $\overline{h} \equiv {\mathbb E}\ h$. The fluctuation in the information $\delta h_i \equiv h_i-\overline{h}$ has the property that it is independent for each observations:
\begin{equation}
{\mathbb E}\ \delta h_i\,\delta h_j = C_0 \delta_{ij}, \label{Eqn:MarkNoise}
\end{equation}
where $C_0$ is a constant and $\delta_{ij}$ is the Kronecker delta, due to the Markovian property. In close analogy to the derivation of AIC, we will Taylor expand the information in terms of the model parameterization ${\bm \theta}$ around the true parameterization ${\bm \theta}_0$. We make the following standard definitions:
\begin{eqnarray}
\delta {\bm \theta} &\equiv& {\bm \theta}-{\bm \theta}_0, \label{againeqn1} \\
\hat{\bm I}_i &\equiv& \nabla_{\bm \theta}\nabla_{\bm \theta}^Th_i(X^N|{\bm \theta}_0), \\
{\bm I} &\equiv& \mathbb{E}_X \nabla_{\bm \theta}\nabla_{\bm \theta}^Th_i(X^N|{\bm \theta}_0), \\
{\bm x}_i &\equiv& \nabla_{\bm \theta}h_i(X^N|{\bm \theta}_0),\\
{\bm X} &\equiv& \sum_{i}{\bm x}_i. \label{againeqn2} 
\end{eqnarray}
where $\delta {\bm \theta}$ is the perturbation in the parameters, ${\bm I}$ and $\hat{\bm I}_i$   are the Fisher Information and its  estimator respectively. The subscript $i$ refers to the $i$th observation. Note that since the true parameterization minimizes the information by definition, ${\mathbb E}\ {\bm x}_i = 0$. Furthermore, Eqn.~\ref{Eqn:MarkNoise} implies that 
\begin{equation}
{\mathbb E}\ {\bm x}_i \, {\bm x}_j^T = {\bm I} \delta_{ij}
\end{equation}
where ${\bm I}$ is the Fisher Information. The Taylor expansion of the information can then be written:
\begin{eqnarray}
h({X}^N|{\bm \theta}) &=& h({X^N}|{\bm \theta}_0)+\delta{\bm \theta}^T{\bm X} + {\textstyle \frac{1}{2}}\delta{\bm \theta}^TN{\bm I} \delta{\bm \theta}+{\cal O}( \delta{\bm \theta}^3), 
\end{eqnarray}
to quadratic order in $\delta{\bm \theta}$.

It is convenient to transform the random variables ${\bm x}_i$ to a new basis in which the Fisher Information is the identity. This is accomplished by the transformation
\begin{eqnarray}
{\bm x}_i' &\equiv& {\bm I}^{-1/2}{\bm x}_i, \\
{\bm \theta}' &\equiv& {\bm I}^{1/2}{\bm\theta}, 
\end{eqnarray}
which results in the following expression for the information:
\begin{eqnarray}
h({\bm \theta}|{X}_I) &=& h(X^N|{\bm \theta}_0)+\delta {{\bm \theta}'}^T{\bm X}' + {\textstyle \frac{1}{2}}N\delta {{\bm \theta}'}^T \delta {\bm \theta}' +{\cal O}(\delta {\bm \theta}^3).
\end{eqnarray}
In our rescaled coordinate system, ${\bm X}'$ can be interpreted as an unbiased random walk of $N$ steps with unit variance in each dimension. 

We  determine the MLE parameter values:
\begin{eqnarray}
\delta {\hat{\bm \theta}}_{X}' = -{\textstyle \frac{1}{N}}{\bm X}'.
\end{eqnarray}
To compute the complexity we need the following expectations of the information:
\begin{eqnarray}
{\mathbb E}_{X,Y}\ h(Y^N|\hat{\bm \theta}_{X}) &=& {\mathbb E}_{X,Y}\ \left\{h(Y^N|{\bm \theta}_0)- {\textstyle \frac{1}{N}}{{\bm X}'}^T{\bm Y}' + {\textstyle \frac{1}{2N}} {{\bm X}'}^2+{\cal O}(\delta {\bm \theta}^3)\right\}, \label{Eqn:SXY} \\
{\mathbb E}_{X}\ h(X^N|\hat{\bm \theta}_{X}) &=& {\mathbb E}_{X,Y}\ \left\{h(X^N|{\bm \theta}_0)- {\textstyle \frac{1}{2N}}{{\bm X}'}^2+{\cal O}(\delta {\bm \theta}^3), \right\}. \label{Eqn:SXX}
\end{eqnarray}
Since the signals $X^N$ and $Y^N$ are independent, the second term on the RHS of Eqn.~\ref{Eqn:SXY} is exactly zero. It is straight forward to demonstrate that 
\begin{equation}
\mathbb{E}_{X} {\bm X}'^2_I = N{\cal d},
\end{equation}
where ${\cal d}$ is the dimension of the parameter $\bm \theta$, which has an intuitive interpretation as the mean squared displacement (${\bm X}'^2$) of a unbiased random walk of $N$ steps in ${\cal d}$ dimensions. The complexity is therefore:
\begin{equation}
{\cal K} \equiv {\mathbb E}_{X,Y}\ \left\{ h(Y^N|\hat{\bm \theta}_{X})-h(X^N|\hat{\bm \theta}_{X}) \right\}=  {\cal d}.
\end{equation}
which is the AIC complexity. To compute the complexity associated with the first binary segmentation, we will compute the nesting complexity ${\cal k}(2)$ using Eqn.~\ref{Eqn:nestingcomplexity}. We will therefore generate the observations $X^N$ and $Y^N$ using the unsegmented model ${\cal M}^1$. Remember that by convention we assign the first change-point index to the first observation $i_1=1$. The optimal but fictitious change-point index for binary segmentation is:
\begin{eqnarray}
\hat{\imath}_2({X}) &=& \arg \min_{1< i \le N} \left\{\right.  h({X}^{[1,i-1]}|\hat{\bm \theta}_{{X}^{[1,i-1]}})+h({X}^{[i,N]}|\hat{\bm \theta}_{{X}^{[i,N]}})\left.\right\},
\end{eqnarray}
where the ${X}^{[j,k]}$ represent the respective partitions of the signal $X^N$ made by the change point $i$. (Note that in the case of an AR process, it is possible to write overlapping partitions to account for the system memory.) The MLE model for two states is defined:
\begin{eqnarray}
\hat{\cal M}_X^2 &\equiv& \left( \begin{array}{cc}
1 & \hat{\imath}_2  \\
\hat{\bm \theta}_{X^{[1,\hat{\imath}_2-1]}}  & \hat{\bm \theta}_{X^{[\hat{\imath}_2,N]}}) \end{array}
 \right).
\end{eqnarray}
To compute the nesting complexity, we compute the difference in the information between the two-state and one-state MLE models:
\begin{eqnarray}
 \nonumber  h({X^N}|\hat{\cal M}^2_X)- h({X^N}|\hat{\cal M}^{1}_X)  &=& \min_{1< i \le N} \left\{\right.  \cancel{h({X}^{[1,i-1]}|{\bm \theta}_0)} +\cancel{h({X}^{[i,N]}|{\bm \theta}_0)} -\cancel{h({X}^{[1,N]}|{\bm \theta}_0)} \\
& & - {\textstyle \frac{1}{2(i-1)}}{{\bm X}'}^2_{[1,i-1]} - {\textstyle \frac{1}{2(N+1-i)}}{{\bm X}'}^2_{[i,N]} + {\textstyle \frac{1}{2N}}{{\bm X}'}^2_{[1,N]} \left.\right\},\label{Eqn:cancel}
\end{eqnarray}
where the terms that are zeroth order in the perturbation cancel since the model is nested and ${\bm X}'_{[i,j]}$ are the ${\bm X}'$ computed in the two partitions of the data. (This equation is analogous to Eqn.~\ref{Eqn:SXX}.) It is straightforward to compute the analogous expression for information difference for signal $Y^N$. The nesting penalty can then be written:
\begin{eqnarray}
{\cal k}_-(2) &\equiv& \underset{q(\cdot|{\cal M}^1_0)}{\mathbb{E}_{{X},{Y}}} \left\{ h(Y^N|\hat{\cal M}_X^2) -h(X^N|\hat{\cal M}_X^2) - h(Y^N|\hat{\cal M}_X^1)  + h(X^N|\hat{\cal M}_X^1) \right\}\\ 
&=& \underset{q(\cdot|{\cal M}^1_0)}{\mathbb{E}_{{X}}} \max_{1< i \le N} \left\{\right. {\textstyle \frac{1}{i-1}}{{\bm X}'}^2_{[1,i-1]}+{\textstyle \frac{1}{N+1-i}}{{\bm X}'}^2_{[i,N]}- {\textstyle \frac{1}{N}}{{\bm X}'}^2_{[1,N]}  \left.\right\}, \label{brid}
\end{eqnarray} 
where the cross terms between signals $X^N$ and $Y^N$ are zero since the signals are independent. It is now convenient to introduce a ${\cal d}$-dimensional discrete Brownian bridge:
\begin{equation}
{\bm B}'_j \equiv {{\bm X}'}_{[1,j]}-{\textstyle \frac{j}{N}}{{\bm X}'}_{[1,N]},
\end{equation}
by using the well known relation between Brownian walks and bridges \cite{WikipediaBridge}. The Brownian bridge has the property that ${\bm B}'_0 = {\bm B}'_N = 0$, where each step has unit variance per dimension and mean zero. After some algebra, the nesting complexity can be written:
\begin{equation}
 {\cal k}_-(2) = \underset{q(\cdot|{\cal M}^1_0)}{\mathbb{E}_{{X}}} \max_{1\le  j < N} \left\{\right. {\textstyle \frac{N}{j(N-j)}}{{\bm B}'}_j^2\left.\right\}. 
\end{equation}
The details of the state model will determine the distribution function for the discrete steps in the Brownian bridge, but the Central Limit Theorem implies that the distribution will approach the normal distribution. Therefore, it is convenient to approximate the discrete Brownian bridge ${\bm B}'_n$ as an idealized Brownian bridge with normally distributed steps:
\begin{equation}
{\bm B}'_j \rightarrow {\bm B}_j \equiv \sum_{i=1}^{n}{\bm b}_i, \ \ {\rm such\ that}\ \ {\bm B}_N = 0, \label{AppCLTapproxEqn2}
\end{equation}
where the ${\bm b}_i$ are steps that are normally distributed with variance one per dimension ${\cal d}$ and mean zero. We now introduce a new random variable $U(N,{\cal d})$, the ${\cal d}$-dimensional Change-Point Statistic \cite{Horvath1993,Horvath1999}:
\begin{equation}
U(N,{\cal d}) \equiv  {\textstyle \frac{1}{2}} \max_{1\le  j < N} {\textstyle \frac{N}{j(N-j)}}  {\bm B}_j^2,\label{Eqn:Ufor1}
\end{equation}
which is a ${\cal d}$-dimensional generalization of the change-point statistic computed by Hawkins \cite{Hawkins1977}.  In terms of the statistic $U$, the nesting penalty is
\begin{equation}
{\cal k}_{-}(2)  = 2\,\mathbb{E}_{U}\ U(N,{\cal d} ) = 2\overline{U}(N,{\cal d} ). 
\end{equation}
We will discuss the connection to the frequentist LPT test shortly.

\begin{figure}
\centering
\includegraphics[width=0.5\textwidth]{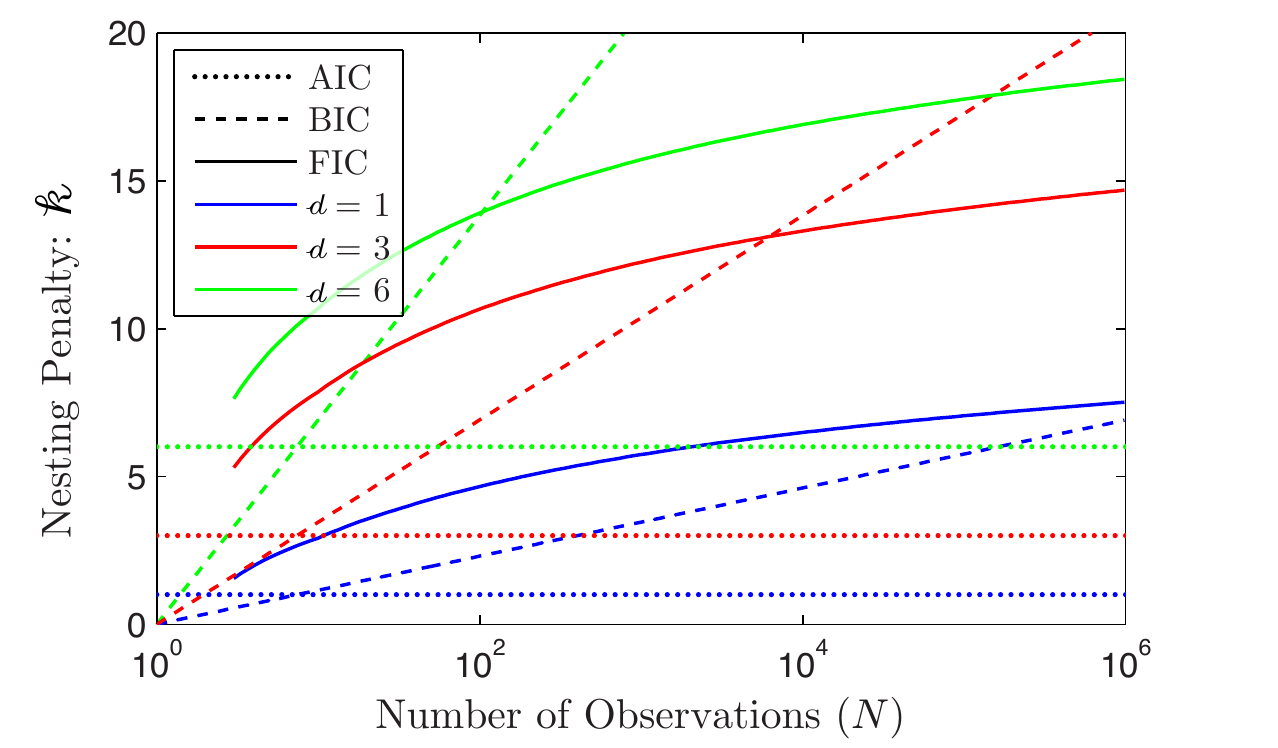}
\caption{{\bf Nesting complexity for AIC, FIC and BIC.} The nesting complexity is plotted for three state dimensions ${\cal d}=\{1,3,6\}$ and $n=2$. First note that the AIC penalty is much smaller than the other two nesting complexities. BIC is empirically known to produce  acceptable results under some circumstances. For sufficiently large samples ($N$), the ${\cal k}_{\rm BIC}>{\cal k}_{\rm FIC}$, resulting in over penalization and the rejection of states that are supported statistically. This effect is more pronounced for large state dimension ${\cal d}$ where the crossover occurs for small observation number $N$. ${\cal k}_{\rm BIC}$ is too small for a wide range of sample sizes, resulting in over segmentation.\label{complexityfig}}
\end{figure}

\medskip
\noindent
{\bf Nesting complexity for $n$ states.} The generalization of the analysis to $n$ states is intuitive and straightforward. In the local binary-segmentation algorithm, segmentation is tested locally. The relevant complexity is computed with respect to the length of the $J$th partition. It is convenient to work with the approximation that all partitions are of equal length since the complexity is slowly varying in $N$. We therefore define the local nesting complexity
\begin{equation}
{\cal k}_{L-}(n)  = 2\,\mathbb{E}_{U}\ U({\textstyle \frac{N}{n-1}},{\cal d} ) = 2\overline{U}({\textstyle \frac{N}{n-1}},{\cal d} ), 
\end{equation}  
where ${\textstyle \frac{N}{n-1}}$ is the mean partition length. The nesting complexity for the binary segmentation of a single state is show in Fig.~\ref{complexityfig} for several different dimensions ${\cal d}$, and compared with the complexity predicted by AIC and BIC.

 In the global binary-segmentation algorithm, the next change-point is chosen by identifying the best position over all intervals. We therefore generalize all our expressions accordingly. We introduce a generalization of the Change-Point Statistic where we replace $N$ with a vector of the lengths of the constituent segment lengths ${\bm N}^n \equiv (N_1,...N_n)$. We now define our new change-point statistic:
\begin{equation}
U_G( {\bm N}^n,{\cal d} ) \equiv \max_{1\le i \le n} U(N_i,{\cal d}). 
\end{equation}
Because it is computationally intensive to compute $U_G$ for all possible segmentations ${\bm N}^n$, we assume that all the partitions are roughly the same size and consider $n$ segments length $N/(n-1)$. Since the complexity is slowly varying in $N$, this does not in general lead to significant information loss. We therefore introduce another change-point statistic:
\begin{eqnarray}
{\cal k}_{G-}(n) &\equiv&  2\,\mathbb{E}_{U} \max_{1\le i \le n}\left\{ \right.U_i({\textstyle \frac{N}{n-1}},{\cal d}) \left. \right\} \quad  \left(\,  \approx 2\,\mathbb{E}_{U} \,  U_G( {\bm N}^n,{\cal d}) \, \right) \end{eqnarray}
that we will apply in the global binary-segmentation algorithm.

\medskip
\noindent
{\bf Series expressions for the nesting complexity.} It is straightforward to compute the asymptotic dependence of the nesting penalty on the number of observations $N$:
\begin{eqnarray}
k_{G-}(n) &\approx& 2 \log \log {\textstyle \frac{N}{n}} + 2 \log n +  {\cal d}  \log \log \log {\textstyle \frac{N}{n}} + ... , \label{kgminus}\\
k_{L-}(n) &\approx& 2 \log \log {\textstyle \frac{N}{n}} +    {\cal d}  \log \log \log {\textstyle \frac{N}{n}} + ... \label{klminus}
\end{eqnarray}
These expression are slowly converging and in practice, we advocate using Monte Carlo integration to determine the nesting penalty. If computationally cumbersome, Eqn. \ref{kgminus} and \ref{klminus} are useful in placing our approach in relation to existing theory. 

Both the local and the global encoding have the same leading-order $2 \log \log N$ dependence that has been advocated by Hannan and Quinn \cite{Hannan1979}, although interestingly not in this context. In contrast, this $2 \log \log N$ dependence is in disagreement with the Bayesian Information Criterion, which has often been applied to change-point analysis. As illustrated by Fig.~\ref{complexityfig}, the BIC complexity:
\begin{equation}
{\cal K}_{\rm BIC} = {\textstyle \frac{\cal d}{2}} \log N,
\end{equation}
can be either too large or too small depending on the number of observations and the dimension of the model. It has long been appreciated that BIC can only be strictly justified in the large-observation-number limit. In this asymptotic limit, the BIC complexity is always larger than the FIC complexity due to the leading order $\log N$ dependence which will tend to lead to under fitting or under segmentation.
It is clear from Fig.~\ref{complexityfig} that large ($N>10^6$) may constitute much larger datasets than are produced in many applications. 

\medskip
\noindent
{\bf Global versus local complexity.} We proposed two possible parameter encoding algorithms above that give rise two distinct complexities: $k_{L-}$ and $k_{G-}$. Which complexity should be applied in the typical problem? For most applications, we expect the number of states $n$ to be proportional to the number of observations $N$. Doubling the length of the dataset will result in the observation of twice as many change points on average. The application of the local nesting complexity clearly has this desired property since it depends on the ratio of $N/n$. It is this complexity we advocate under most circumstances.

In contrast the global nesting complexity contains an extra contributions to the complexity $2 \log n$. The reason is intuitive: In the global binary segmentation algorithm, one picks the best change point among $n$ segments and therefore complexity must reflect this added degree of choice. Consequently  a larger feature must be observed to be above the expected background. The use of the global nesting complexity makes a statement of statistical significance against the entire signal, not just against a local region. In the context of discussing the significance of the observation of a rare state that occurs just once in a dataset, the global nesting complexity is the most natural metric of significance.

\begin{figure}
\centering
\includegraphics[width=1.0\textwidth]{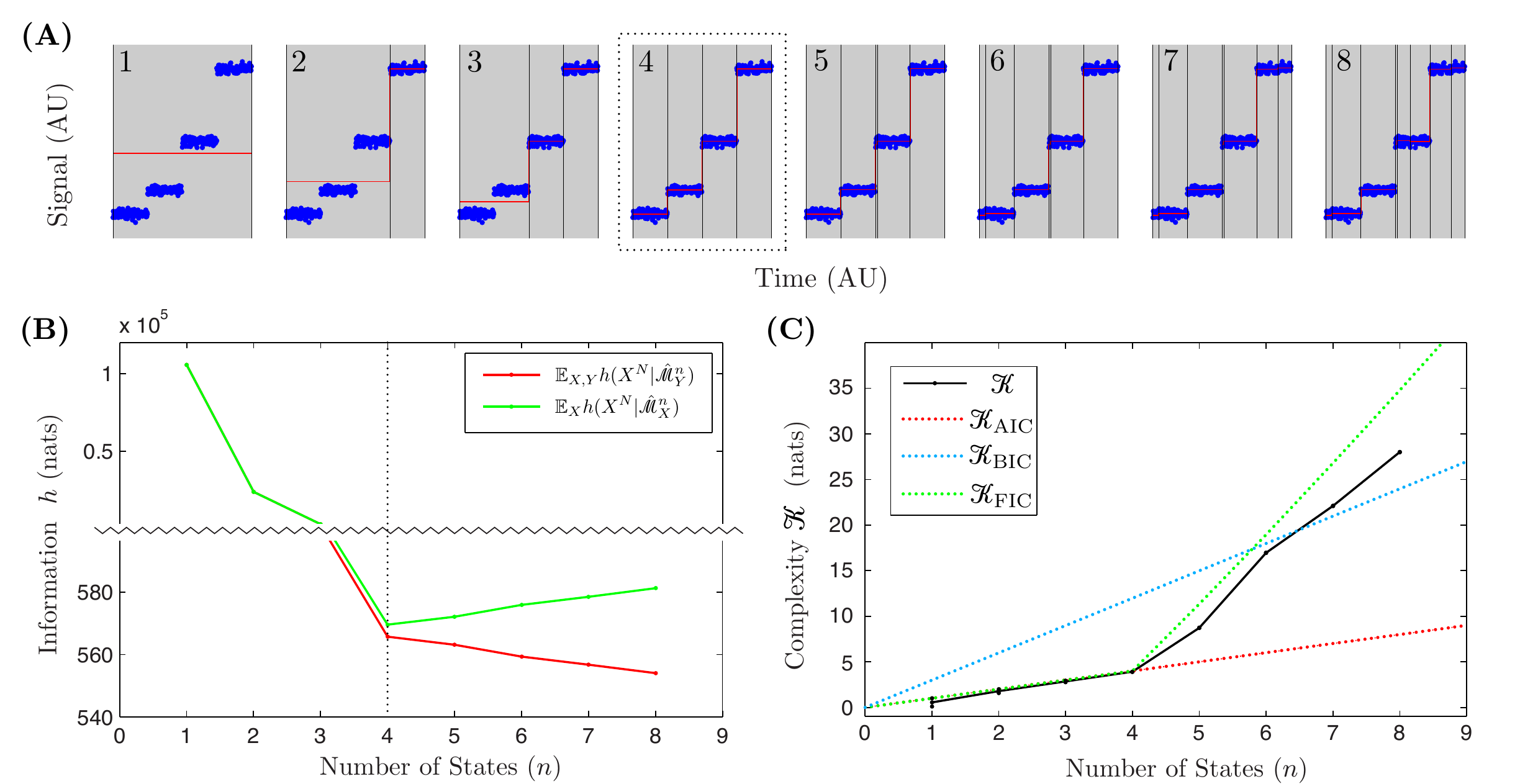}
\caption{{\bf Information-based model selection. Panel A: Nested models generated by a Change-Point Algorithm.} 
Simulated data (blue points) generated by a true model with four states is fit to a family of nested models (red lines) using a Change-Point Algorithm. Models fit with $1\le n \le 8$ states are plotted. The fit change points are represented as vertical black lines. The number of states ($n$) in each fit model is shown in the top-left corner of each panel. The true model has four states and the fit model with four states is indicated with a dotted box. The models with five through eight states have superfluous states that are not present in the true model. 
{\bf Panel B: Four changes points minimizes information loss.} Both the expectation of the information  (red) and the cross entropy (green) are plotted as a function of the number of states $n$. The y-axis ($h$, information) is split to show the initial large changes in $h$ as well as the subsequent smaller changes for  $4\le n \le 8$. The cross entropy (green) is minimized by the the model that best approximates the truth ($n=4$). The addition of parameters leads to an increase in cross entropy (less predictive) as a consequence of the addition of superfluous parameters, as indicated by the increase of the cross entropy (green) for $n\ge4$. The information loss estimator (red) is biased and continues to decreases with the addition of states as a consequence of over fitting. In an experimental context only the information can be computed since the true distribution is unknown. {\bf Panel C: Complexity of Change-Point Analysis.} The true complexity is computed for the model shown in panel A via Monte Carlo simulation for $10^6$ realizations of the observations $X^N$ and compared with three models for the complexity AIC, FIC and BIC. For models with states numbering  $1\le n\le 4$, the true complexity (black) is correctly estimated by the AIC complexity (red dotted) and the FIC complexity (green). But for a larger number of states ($4\le n \le 8)$, only FIC accurately estimates the true complexity.\label{modelselection}}
\end{figure}

\medskip
\noindent
{\bf Computing the complexity from the nesting complexity.} To compute the FIC complexity, we sum the nesting complexities using Eqn.~\ref{Eqn:complexity}. For datasets with identifiable change points, the FIC complexity is initially identical to AIC:
\begin{equation}
{\cal K}_{\rm FIC}(n) = n{\cal d}, 
\end{equation}
until the change in the information on nesting $\Delta h<{\cal k}_-$, when FIC predicts that there is a change in slope in the penalty. The FIC, AIC, and BIC predicted complexities are compared with the true complexity for an explicit change-point analysis in Fig.~\ref{modelselection}, Panel C. It is immediately clear from this example that FIC quantitatively captures the true dependence of the penalty, including the change in slope at $n=4$, exactly as predicted by the FIC complexity. As predicted, the AIC complexity is initially correct until the segmentation process must be terminated. At this point the complexity increases significantly with the result that the AIC complexity fails to terminate the segmentation process. In contrast, the BIC complexity is initially too large, but fails to grow at a sufficient pace to match the true complexity for $n>4$.

\section{The relation between frequentist and information-based approach}
\label{AppLPTSec}
Consider the LPT test for the following problem: We propose the binary segmentation of a single partition. In the null hypothesis ($H_0$) is the partition is described by a single state (unknown model parameters ${\bm \theta}_0$) and the hypothesis to be tested ($H_1$) is that the partition is actually sub-divided into two states (unknown change point and model parameters ${\bm \theta}_1$ and ${\bm \theta}_2$). We  use the log-likelihood ratio as the test statistic:
\begin{equation}
V(X^N) \equiv \log \frac{q(X^N|\hat{\cal M}^2_X)}{ q(X^N |\hat{\cal M}^1_X)} =  h(X^N |\hat{\cal M}^1_X)-h(X^N|\hat{\cal M}^2_X).
\end{equation}
In the Neyman-Pearson approach to hypothesis testing, we assume the null hypothesis (1 state) and compute the distribution in the test statistic $V$. As before, we will expand the information around the true parameter values ${\bm \theta}_0$. In exact analogy to Eqn.~\ref{Eqn:cancel}, we find that $V$ and our previously defined statistic $U$ identically distributed: 
\begin{equation}
V \sim U,
\end{equation}
up to the approximations discussed in the derivation. Therefore we will simply refer to $V$ as $U$.

In the canonical frequentist approach we specify a critical test statistic value $u_\gamma$ above which the alternative hypothesis is accepted. $u_\gamma$ is selected such that the alternative hypothesis $H_1$ is rejected given that the null hypothesis $H_0$ is true with a probability equal to the confidence level $\gamma$: 
\begin{equation}
\gamma = F_U(u_\gamma),
\end{equation} 
where $F_U$ is the cumulative distribution of $U$. 

Therefore  we can interpret both the information-based approach and the frequentist approach as making use of the same statistic $U$. In the frequentist approach, a confidence level ($\gamma$) is specified to determine the critical value $u_\gamma$ with which to accept the two-state  hypothesis. The information-based approach also uses the statistic $U$, but the critical value of the statistic (${\cal k}_-$) is computed from the distribution of the statistic itself ${\cal k}_-=2\overline{U}$. The information-based approach chooses the confidence level that optimizes predictivity.

\section{Applications}
In the interest of brevity we have not included analysis of either experimental data or simulated data with a signal-model dimension larger than one, but we have tested the approach extensively. For instance, we have applied this technique to an experimental single-molecule biophysics application that is modeled by an Ornstein-Uhlenbeck process with state-model dimension of four \cite{CPshort}. We also applied the approach in other  biophysical contexts including the analysis of bleaching curves, cell and molecular-motor motility \cite{CPApp}.

\section{Discussion}

In this paper, we present an information-based approach to change-point analysis using the Frequentist Information Criterion (FIC). The information-based approach to inference provides a powerful framework in which models with different parameterization, including different model dimension, can be compared to determine the most predictive model. The model with the smallest information criterion has the best expected predictive performance against a new dataset.

Our approach has two advantages over existing frequentist-based ratio tests for change-point analysis: (i) We derive an FIC complexity that depends only on the dimension of the state  model (${\cal d}$), the number of states ($n$) and observations ($N$). Therefore it may be unnecessary to develop and compute custom statistics for specific applications. (ii) In the frequentist approach one must specify an {\it ad hoc} confidence level to perform the analysis. In the information-based approach, the confidence level is chosen automatically based upon the model complexity. The information-based approach is therefore parameter and prior free. 

As the number of change-points increases, the model complexity is observed to transition between an AIC-like  complexity ${\cal O}(N^0)$ and a Hannan-and-Quinn-like complexity ${\cal O}(\log \log N)$. We propose an approximate piecewise expression for this transition.   The computation of this approximate model complexity can be interpreted as the expectation of the extremum of a ${\cal d}$-dimensional Brownian bridge. We believe this information-based approach to change-point analysis will be widely applicable.

\subsection*{Author Contributions}
P.A.W. and C.H.L.~designed research; performed research; contributed analytic tools; analyzed data; or wrote the paper.

\acknowledgements{P.A.W.~and C.H.L.~would like to thank  K.~Burnham, J.~Wellner, L.~Weihs and M.~Drton for advice and discussions, D.~Dunlap and L.~Finzi for experimental data and M.~Lind\'en and N.~Kuwada for advice on the manuscript. This work was supported by NSF MCB grant 1243492. }


\bibliography{ModelSelection}

\pagebreak
\appendix

\begin{table}
\begin{center}
\framebox{
\parbox{18cm}{ \begin{center}
\parbox{17cm}{
\begin{center}
{\bf Global Binary-Segmentation Algorithm}
\end{center}
\begin{enumerate}
\item{Initialize the change-point vector: ${\bm i} \gets \{1\}$}
\item{Segment model $\hat{\cal M}({\bm i})$:}
\begin{enumerate}
\item{ Compute the entropy change that results from all possible new change-point indices $j$: 
\begin{equation}
\Delta h_j \gets \hat{h}(\{i_1,...,j,...,i_{n}\}|{X})-\hat{h}({\bm i}|{X}), 
\end{equation}}
\item{ Find the minimum information change $\Delta h_{\rm min}$, and the corresponding index $j_{\rm min}$.}
\item{ {\bf If} the information change plus the nesting complexity is less than zero:
\begin{equation}
\Delta h_{\rm min}+{\cal k}_{G-}<0
\end{equation}
{\bf then} accept the change-point $j_{\rm min}$}
\begin{enumerate}
\item{Add the new change-point to the change-point vector.
\begin{equation}
{\bm i} \gets \{i_1,...,j_{\rm min},...i_{n+1}\}
\end{equation}}
\item{Segment model $\hat{\cal M}({\bm i})$} 
\end{enumerate}
\item {{\bf Else} terminate the segmentation process.}
\end{enumerate}
\end{enumerate}
}
\end{center}
}}
\end{center}
\caption{ A global algorithm for binary segmentation. The information $\hat{h}$ is implicitly evaluated at the MLE state-model parameters $\hat{\bm \Theta}$. \label{Tab:Global}  }
\end{table}

\begin{table}
\begin{center}
\framebox{
\parbox{18cm}{ \begin{center}
\parbox{17cm}{
\begin{center}
{\bf Local Binary-Segmentation Algorithm}
\end{center}
\begin{enumerate}
\item{Initialize the change-point vector: ${\bm i} \gets \{1\}$, $I \gets 1$}.
\item{Segment model $\hat{\cal M}({\bm i})$ on state $I$:}
\begin{enumerate}
\item{ Compute the entropy change that results from all possible new change-point indices $j$ on the interval $[i_I,i_{I+1})$: 
\begin{equation}
\Delta h_j \gets \hat{h}(\{...i_I,j,i_{I+1},...\}|{X})-\hat{h}({\bm i}|{X}), 
\end{equation}}
\item{ Find the minimum information change $\Delta h_{\rm min}$, and the corresponding index $j_{\rm min}$.}
\item{ {\bf If} the information change plus the nesting complexity is less than zero:
\begin{equation}
\Delta h_{\rm min}+{\cal k}_{-{\rm L}}<0
\end{equation}
{\bf then} accept the change-point $j_{\rm min}$}
\begin{enumerate}
\item{Add the new change-point to the change-point vector.
\begin{equation}
{\bm i} \gets \{...,i_I,j_{\rm min},i_{I+1},...\}
\end{equation}}
\item{Segment model $\hat{\cal M}({\bm i})$ on states $I$ and $I+1$. }
\item{Merge the resulting index lists.} 
\end{enumerate}
\item {{\bf Else} terminate the segmentation process.}
\end{enumerate}
\end{enumerate}
}
\end{center}
}}
\end{center}
\caption{ A local algorithm for binary segmentation. The information $\hat{h}$ is implicitly evaluated at the MLE state model parameters $\hat{\bm \Theta}$. \label{Tab:Local}  }
\end{table}

\subsection{Type I errors (false positives)}
\begin{figure}
\centering
\includegraphics[width=0.60\textwidth]{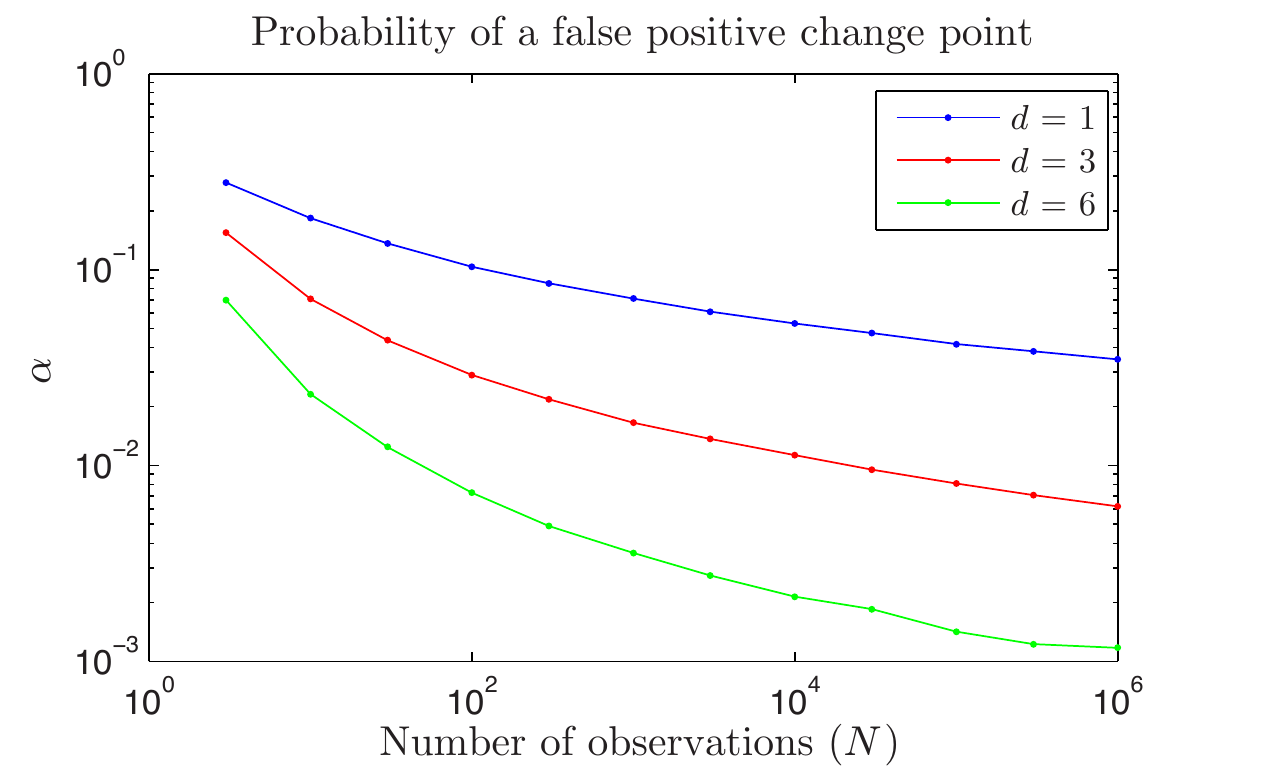}
\caption{{\bf Probability of a false positive change-point.} The probability of a false positive change-point is shown as a function of the number of observations in the interval length $N$ for three different model dimensions.
\label{false_positive_fig}}
\end{figure}

In terms of the Cummulative Probability Distribution (CDF), the probability of a false positive change-point is:
\begin{equation}
\alpha =  1-F_{U}(2\overline{U}),
\end{equation}
where $U$ is the relevant change-point statistic and $\overline{U}$ is its expectation. Using the local binary-segmentation algorithm, $\alpha$ corresponds to the probability of a false positive per data partition and the change-point statistic is defined by Eqn.~\ref{Eqn:Ufor1} evaluated at the average partition length $N_p \equiv {\textstyle \frac{N}{n}}$. The false positive change-point acceptance probability is plotted in Figure \ref{false_positive_fig}.

The analogous false positive rate for the global binary-segmentation algorithm describes the probability of a false positive in the entire data set, including all partitions. In this cases, we use the change-point statistic defined by Eqn.~\ref{Eqn:Ufor2}.

\subsection{ Asymptotic form of the complexity function}
\label{AsympAppSec}
In order to discuss the scaling of the complexity relative to the BIC complexity, we need to derive an asymptotic form for the complexity in the large $N$ limit. We do not recommend explicitly using this asymptotic expression for the complexity for Change-Point Analysis since it converges to the true complexity very slowly, especially for large ${\cal d}$. 

\begin{figure}
\centering
\includegraphics[width=1.0\textwidth]{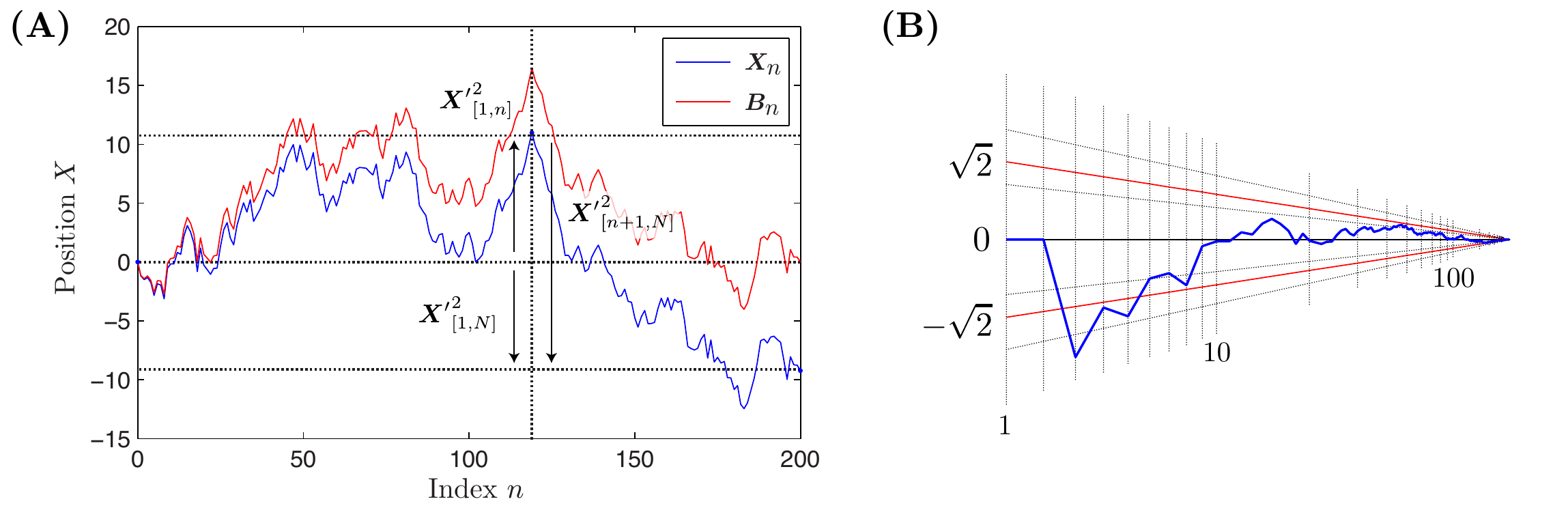}
\caption{{\bf  Panel A:  Brownian Walk and Brownian Bridge.}  A visualization of a random walk ${\bm X}'_[1,n]$ (blue) and the corresponding Brownian bridge ${\bm B}'_n$ (red). {\bf Panel B: Law of Iterated Logs.} A visualization of $S_n/\sqrt{n\log \log n}$ (blue) plotted as an orthographic projection as a function of $n$. $\sqrt{2}$ (red) is the limit of the supremum. }
\label{simtest}
\end{figure}

First let us consider related results for and Brownian walk rather than a Brownian bridge. Let us define $S_n$ as follows:
\begin{eqnarray}
S_n &\equiv& \left|{\bm Z}_{n}\right| \\ 
{\bm Z}_{n} &\equiv& \sum^n_{i=1} {\bm z}_i
\end{eqnarray}
where the ${\bm z}_i$ are independent normally-distributed random variables with mean zero variance one per dimension ${\cal d}$. The Law of Iterated Logs states that \cite{Khinchine1924,Kolmogoroff1929,LILwiki}:
\begin{equation}
\limsup_{n\rightarrow \infty} \frac{S_n}{\sqrt{ n\log \log n}} = \sqrt{2}\ \ \ {\rm a.s.},
\end{equation}
where a.s. is the acronym for almost surely. (See Figure \ref{simtest}.) This behavior of $S_n$ is described in more detail by the  Darling-Erd\"os Theorem \cite{Darling1956}. Let us define a new random variable
\begin{equation}
U'(N_p) \equiv \max_{1\le n \le N_p} \frac{S_n}{\sqrt{n}}, 
\end{equation}
in ${\cal d}=1$ dimensions, the asymptotic cumulative distribution of $U'$ approaches the cumulative distribution for a Gumbel Distribution \cite{Darling1956}:
\begin{eqnarray}
\lim_{N_p\rightarrow \infty} {\rm Pr}\left[ U' < \beta t + u \right] &=& \exp\left[-e^{-t}\right], \\
\beta(N_p) &\equiv& \left( 2\log \log N_p \right)^{-1/2},\\ 
u(N_p) &\equiv& \beta \left[\beta^{-2} + {\textstyle \frac{1}{2}}  \log \log \log N_p - \log 2 \pi^{1/2}\right],
\end{eqnarray}
where $\rm Pr$ denotes probability and the distribution parameters $u$ and $\beta$ are called the location and scale respectively and the average partition length is $N_p \equiv {\textstyle \frac{N}{n}}$.
Let us introduce the cumulative distribution function for $U$:
\begin{equation}
F_U (U) \equiv  {\rm Pr}\left[ U'_{(n)} < U \right].
\end{equation}
This expression can be reordered to put it in the canonical form of the Gumbel Distribution \cite{WikipediaGumbel}: 
\begin{eqnarray}
F_U (U)  &=& \exp\left[-\exp\left(-{\textstyle\frac{U-u}{\beta}}\right)\right], 
\end{eqnarray}
We can then use the well known expression in terms the cdf to compute the cdf of the maximum of $n$ random variables $U'$: 
\begin{eqnarray}
{\rm Pr}\left[ U'_{(n)} < U_{(n)} \right] &=& F_U^n(U),\\
&=& \left(\exp\left[-\exp\left(-{\textstyle\frac{U-u}{\beta}}\right)\right]\right)^n, \\
  &=& \exp\left[-\exp\left(-{\textstyle\frac{U-u_n}{\beta}}\right)\right], 
\end{eqnarray}
where 
\begin{equation}
u_n \equiv u + \beta \log n.
\end{equation}
The mean and variance of the Gumbel Distribution are well known, allowing us to compute the expectation of ${U'}^2_{(n)}$:
\begin{eqnarray}
\mathbb{E}_{x}\ {U'}^2_{(n)} &\approx& \left( u_n+\cancel{\beta \gamma}\right)^2 + \cancel{\frac{\pi^2}{6}\beta^2},\\
&\approx& 2 \log \log N_p + 2 \log n + ... 
\end{eqnarray}
where $\gamma$ is the  Euler-Mascheroni constant and we have used the cancel notation to show which terms have been dropped to lowest order. In the second line, we have written the expression to lowest order in $N$ and $n$.

Horv\'ath has generalized the  Darling-Erd\"os Theorem for a Brownian bridge in ${\cal d}$ dimensions for the application to Change-Point Analysis in the context of the LPT test \cite{Horvath1993,Horvath1999}. The generalized expression for the cumulative distribution leads to a change in the expression for $u$ only:
\begin{equation}
u_{\cal d}(N_p) \equiv \beta \left[\beta^{-2} + {\textstyle \frac{\cal d}{2}}  \log \log \log N_p - \cancel{\log \Gamma\left({\textstyle\frac{\cal d}{2}}\right) }\right]
\end{equation}
where $\Gamma$ is the Gamma Function. We drop the last term since it is not leading order for large $N_p$. We now follow the same steps to generate the distribution for the maximum of $n$ random variables $U'$, leading to a new Gumbel Distribution with location $\mu_{n,{\cal d}}$:
\begin{equation}
u_{n,{\cal d}}(N_p) = \beta \left[\beta^{-2} + {\textstyle \frac{\cal d}{2}}  \log \log \log N_p +\log n\right]
\end{equation}
We now recompute the expectation for ${\cal d}$ dimensions:
\begin{eqnarray}
{\cal k}_{-}^{\rm G}(N_p,n,{\cal d}) &\equiv& \mathbb{E}_{x}\ {U'}^2_{(n)}(N_p,n,{\cal d}), \\
  &\approx& \left( u_{n,{\cal d}}+\beta \gamma\right)^2 + \frac{\pi^2}{6}\beta^2\\
&\approx& 2 \log \log N_p + 2 \log n +  {\cal d}  \log \log \log N_p + ... 
\end{eqnarray}
where we have kept terms only to highest order in $n$ and $N_p$.

\end{document}